# Sub-micrometer near-field focusing of spin waves in ultrathin YIG films

B. Divinskiy[1*], N. Thiery[2], L. Vila[2], O. Klein[2], N. Beaulieu[3], J. Ben Youssef[3], S. O. Demokritov[1], and V. E. Demidov[1]

[1]*Institute for Applied Physics and Center for NanoTechnology, University of Muenster, 48149 Muenster, Germany*

[2]*Univ. Grenoble Alpes, CNRS, CEA, Grenoble INP, IRIG-SPINTEC, F-38000 Grenoble, France*

[3]*LabSTICC, CNRS, Université de Bretagne Occidentale, 29238 Brest, France*

We experimentally demonstrate tight focusing of a spin wave beam excited in extended nanometer-thick films of Yttrium Iron Garnet by a simple microscopic antenna functioning as a single-slit near-field lens. We show that the focal distance and the minimum transverse width of the focal spot can be controlled in a broad range by varying the frequency/wavelength of spin waves and the antenna geometry. The experimental data are in good agreement with the results of numerical simulations. Our findings provide a simple solution for implementation of magnonic nano-devices requiring local concentration of the spin-wave energy.

*Corresponding author, e-mail: b_divi01@uni-muenster.de

The advent of high-quality nanometer-thick films of magnetic insulator Yttrium Iron Garnet (YIG)[1-3] essentially expanded horizons for the field of magnonics[4-6] utilizing spin waves for transmission and processing of information on the nanoscale. Thanks to the small thickness and ultra-low magnetic damping, these films enable implementation of magnonic devices with nanometer dimensions,[7-8] where the spin-wave losses are by several orders of magnitude smaller compared to those in devices based on metallic ferromagnetic films.[9] The large propagation length of spin waves in YIG is particularly beneficial for implementation of spatial manipulation of spin-wave beams in the real space.

It is now well established that the propagation of spin waves can be controlled by using approaches similar to those used in optics.[10-16] However, in contrast to light waves, the wavelength of spin waves can be as small as few tens of nanometers,[17,18] which allows one to implement efficient wave manipulation on the nanoscale. In the recent years, particular attention was given to the possibility to controllably focus propagating spin waves.[10-14,19,20] Such focusing allows one to concentrate the spin-wave energy in a small spatial area, which is important, for example, for implementation of the efficient local detection of spin-wave signals. Provided that the position of focal point is controllable by the spin-wave frequency, the focusing can also be utilized for implementation of the frequency multiplexing.[21] Additionally, the strong local concentration of the spin-wave energy can be used to stimulate nonlinear phenomena, for example, the second-harmonic generation.[22]

Efficient spin-wave focusing can be achieved relatively easily in confined geometries, such as stripe waveguides,[23] where it is governed by the interference of multiple co-propagating quantized spin-wave modes.[24] In the case of extended magnetic films, implementation of the focusing appears to be less straightforward. In the recent years, several approaches have been suggested utilizing a spatial variation of the effective spin-wave refraction index,[14,16] refraction of spin waves at the modulation of the film thickness[11] or the temperature,[12] diffraction from a Fresnel zone plate,[13] and excitation of spin-wave beams by curved transducers.[19,20] All these approaches are rather complex in terms of practical implementation, particularly on the nanoscale. A much simpler approach known in optics[25-27] relies on the utilization of Fresnel diffraction patterns appearing in the near-field region of a single slit, where the Fresnel number $F=a^2/(\lambda d)$ is of the order or larger than 1 (Ref. 28). Here $a$ is the length of the slit, $\lambda$ is the wavelength, and $d$ is the distance from the slit to the observation point. As was experimentally shown for light waves[26] and surface plasmon polaritons,[27] such a slit functions as an efficient near-field lens enabling tight focusing of the incident wave.

In this work, we demonstrate experimentally that the principles of near-field diffractive focusing are also applicable for spin waves in in-plane magnetized magnetic films, which, in contrast to light, exhibit anisotropic dispersion. By using a 2 μm long spin-wave antenna, which is equivalent to a single one-dimensional slit[29], we achieve focusing of the excited spin waves into an area with the transverse width below 700 nm. We show that, in agreement with the theory developed for light waves, the focal distance increases with the decrease of the wavelength of the spin wave, which allows electronic control of the spin-wave focusing by the frequency and/or the static magnetic field. We also perform micromagnetic simulations, which show excellent agreement with the experimental data and allow us to analyze the effects of the antenna geometry on the focusing characteristics.

Figure 1(a) shows the schematic of the experiment. The test devices are based on a 56 nm thick YIG film grown by liquid phase epitaxy on a gadolinium gallium substrate. The independently determined saturation magnetization of the film is $4\pi M$=1.78 kG and the Gilbert damping parameter is $\alpha=1.4\times10^{-4}$. The YIG film is magnetized to saturation by the static magnetic field $H$ applied in the film plane. The excitation of spin waves is performed by using lithographically defined 2 μm long, 300 nm wide, and 7 nm thick spin-wave antenna contacted by 2 μm wide and 30 nm thick Au microstrip lines[30]. The microwave-frequency electrical current $I_{MW}$ flowing through the antenna creates a dynamic magnetic field $h$, which couples to the magnetization in the YIG film and excites spin waves propagating away from the antenna. Figure 1(b) shows the normalized spatial distribution of the amplitude of the dynamic field created by the antenna in the YIG film calculated by using COMSOL Multiphysics simulation software (https://www.comsol.com/comsol-multiphysics). As seen from these data, due to the large difference in the width of the antenna and the microstrip lines, the amplitude of the dynamic field underneath the antenna is by an order of magnitude larger compared to that underneath the lines. Therefore, the efficient spin wave excitation is only possible in the 2 μm long antenna region. This disbalance is further enhanced for spin waves with wavelengths comparable or smaller than the width of the microstrip lines, due to the reduced coupling efficiency of the inductive mechanism.[9]

Spatially resolved detection of excited spin waves is performed by micro-focus Brillouin light scattering (BLS) spectroscopy.[9] The probing light with the wavelength of 473 nm and the power of 0.1 mW produced by a single-frequency laser is focused into diffraction-limited spot on the surface of the YIG film. By analyzing the spectrum of light inelastically scattered from magnetic excitations, we obtain signal – the BLS

intensity – proportional to the intensity of spin waves at the location of the probing spot. By scanning the spot over the sample surface, we obtain spatial maps of the spin-wave intensity. Additionally, by using the interference of the scattered light with the reference light modulated at the excitation frequency,[9] we record spatial maps of the spin-wave phase.

Figures 2(a) and 2(b) show the representative intensity and phase maps recorded at $H$=500 Oe by applying excitation current with the frequency $f$=3.8 GHz. The power of the applied signal is 10 µW, which is proven to provide the linear regime of excitation and propagation of spin waves. In agreement with the above discussion, spin waves are only radiated from the region of the narrow antenna. More importantly, the radiated beam exhibits significant narrowing and increase of the intensity at the distance $d$=3.6 µm from the center of the antenna clearly indicating focusing of the excited spin waves. Qualitatively similar behaviors were also observed for different excitation frequencies in the range $f$=3.2-4 GHz, although the distance $d$ was found to change strongly with the variation of $f$.

From the phase-resolved measurements (Fig. 2(b)) we obtain the wavelength $\lambda$=0.6 µm of spin waves at $f$=3.8 GHz. By repeating these measurements for different excitation frequencies, we obtain the spin-wave dispersion curve (Fig. 2(c)), which allows us to relate the excitation frequency to the spin-wave wavelength. Note that the experimental data (symbols in Fig. 2(c)) are in perfect agreement with the results of calculations (curve in Fig. 2(c)) based on the analytical theory (Ref. 31).

On one side, the observed focusing is counterintuitive. Indeed, the excitation of waves by a finite-length straight antenna, as used in our experiment, is equivalent to a diffraction of a wave with an infinite plane front from a slit[29], which is known to result in a formation of a divergent beam. On the other side, it is also known[25-27] that, before the beam starts to diverge, a complex focusing-like diffraction pattern is formed in the near field just behind a slit. In the recent years, it was shown theoretically[25] and proven experimentally,[26,27] that these near-field effects can be used for efficient focusing of waves of different nature.

Due to the insufficient spatial resolution, the fine details of the near-field spin-wave pattern cannot be seen in the experimental maps (Fig. 2(b)). Therefore, we perform micromagnetic simulations using the software package MuMax3 (Ref. 32). We consider a magnetic film with dimensions of 20 µm ×10 µm ×0.05 µm discretized into 10 nm ×10 nm ×50 nm cells. The standard for YIG exchange constant of 3.66 pJ/m is used. The spin waves are excited by applying a sinusoidal dynamic magnetic field with the amplitude of 1 Oe, which is close to the estimated experimental value of 3 Oe. The spatial distribution of the excitation field is taken from COMSOL simulations (Fig. 1(b)). The angle of the excited magnetization precession is of the order of 0.1°.

The results of simulations for the excitation frequency $f$=3.8 GHz and $H$=500 Oe are shown in Fig. 3(a). The simulated map of the normalized spin-wave intensity $<M_x^2>/<M_x^{2\,max}>$ exhibits a narrowing of the excited beam and concentration of the spin-wave energy in exactly the same way, as it is observed in the experiment (compare with Fig. 2(a)). Simultaneously, it shows a fine structure, which is reminiscent of that obtained for light diffracted on a slit.[26] To further confirm the analogy, we perform simulations for the case of plane spin waves diffracting from a slit formed by two 300 nm wide rectangular regions with increased magnetic damping ($\alpha$=1) with a 2 µm long gap between them (Fig. 3(b)). The close similarity between the obtained patterns shows that the experimental results obtained for spin-wave excitation by the antenna are equally applicable for spin-wave focusing by a slit lens. We emphasize that such a lens can be easily implemented in practice, for example, by using ion implantation into nanometer-thick YIG film.

To additionally address the effects of the anisotropy of the spin-wave dispersion, we show in Fig. 3(c) an intensity map calculated for spin waves in an out-of-plane magnetized film characterized by an isotropic dispersion[33]. Comparison of Fig. 3(c) with Fig. 3(a) shows that the anisotropy makes the near-field focusing even better pronounced due to the existence of the preferential direction of the energy flow characterized by the angle θ=17° (see Fig. 3(a)) calculated according to Ref. 34.

Figure 4 shows the quantitative comparison of the experimental results with those obtained from simulations. In Fig. 4(a), we plot one-dimensional sections of the experimental (Fig. 2(a)) and the calculated (Fig. 3(a)) intensity maps along the axis of the spin-wave beam at $z=0$. Both data sets show perfect agreement. In both cases, the intensity increases during the first 3 μm of propagation and reaches a maximum at the distance $y\approx 3.6$ μm, which can be treated as a focal distance. Transverse sections of the intensity maps at this distance (Fig. 4(b)) also exhibit very similar narrowing of the beam to 670±20 nm. As seen from Figs. 4(c) and 4(d), the good agreement between the experimental data and the result of simulations is observed in a broad range of spin-wave wavelengths.

We note that, in contrast to the far-field focusing, in our case, the focal distance depends strongly on the wavelength (Fig. 4(c)). This dependence is in agreement with the theory of the near-field diffractive focusing of light,[25] which predicts that the focal distance should increase with the decrease of the wavelength. This dependence can be particularly important for magnonic applications, since it allows one to focus spin waves with different frequencies at different spatial locations. Alternatively, the focal position can be controlled by the variation of the static magnetic field at the fixed spin-wave frequency.

As seen from Fig. 4(d), the transverse width $w$ of the spin-wave beam at the focal position exhibits a monotonous decrease with the decrease of the wavelength $\lambda$. Therefore, similarly to the far-field focusing, one can obtain stronger concentration of the energy for spin waves with smaller wavelengths. Note, however, that the ratio $w/\lambda$ increases at smaller $\lambda$, making the focusing of short-wavelength spin waves less efficient.

To study the effects of the antenna geometry on the focusing efficiency, we perform micromagnetic simulations for different lengths of the spin-wave antenna $a$ at the fixed value of the wavelength $\lambda=0.6$ μm. The results of these simulations show (Fig. 5) that the focal-point width $w$ generally reduces with decreasing $a$. Therefore, more tight focusing can be achieved by using smaller antennae or slit lenses. Additionally, as can be seen from Fig. 5, the reduction of the antenna length $a$ results in a decrease of the focal distance, which allows one to achieve stronger focusing in devices with smaller dimensions.

In conclusion, we have experimentally demonstrated a simple and efficient approach to the focusing of spin waves on the sub-micrometer scale. The obtained results are applicable not only to the excitation-stage focusing, but can also be used for implementation of near-field lenses for plane spin waves. Our findings significantly simplify the implementation of nano-magnonic devices utilizing spin-wave focusing, which is critically important for their real-world applications.

We acknowledge support from Deutsche Forschungsgemeinschaft (Project No. 423113162) and the French ANR Maestro (No.18-CE24-0021).

**References**

1. Y. Sun, Y. Y. Song, H. Chang, M. Kabatek, M. Jantz, W. Schneider, M. Wu, H. Schultheiss, and A. Hoffmann, Appl. Phys. Lett. **101**, 152405 (2012).

2. O. d'Allivy Kelly, A. Anane, R. Bernard, J. Ben Youssef, C. Hahn, A. H. Molpeceres, C. Carretero, E. Jacquet, C. Deranlot, P. Bortolotti, R. Lebourgeois, J.-C. Mage, G. de Loubens, O. Klein, V. Cros, and A. Fert, Appl. Phys. Lett. **103**, 082408 (2013).

3. C. Hauser, T. Richter, N. Homonnay, C. Eisenschmidt, M. Qaid, H. Deniz, D. Hesse, M. Sawicki, S. G. Ebbinghaus, and G. Schmidt, Sci. Rep. **6**, 20827 (2016).

4. S. Neusser and D. Grundler, Adv. Mater. **21**, 2927 (2009).

5. V. V. Kruglyak, S. O. Demokritov, and D. Grundler, J. Phys. D: Appl. Phys. **43**, 264001 (2010).

6. A.V. Chumak, V.I. Vasyuchka, A.A. Serga, and B. Hillebrands, Nat. Phys. **11**, 453 (2015).

7. S. Li, W. Zhang, J. Ding, J. E. Pearson, V. Novosad, and A. Hoffmann, Nanoscale **8**, 388 (2016).

8. Q. Wang, B. Heinz, R. Verba, M. Kewenig, P. Pirro, M. Schneider, T. Meyer, B. Lägel, C. Dubs, T. Brächer, and A. V. Chumak, Phys. Rev. Lett. **122**, 247202 (2019).

9. V. E. Demidov and S. O. Demokritov, IEEE Trans. Mag. **51**, 0800215 (2015).

10. G. Csaba, A. Papp, and W. Porod, J. Appl. Phys. **115**, 17C741 (2014).

11. J.-N. Toedt, M. Mundkowski, D. Heitmann, S. Mendach, and W. Hansen, Sci. Rep. **6**, 33169 (2016).

12. O. Dzyapko, I. V. Borisenko, V. E. Demidov, W. Pernice, and S. O. Demokritov, Appl. Phys. Lett. **109**, 232407 (2016).

13. J. Gräfe, M. Decker, K. Keskinbora, M. Noske, P. Gawronski, H. Stoll, C. H. Back, E. J. Goering, and G. Schütz, arXiv:1707.03664 (2017).

14. N. J. Whitehead , S. A. R. Horsley, T. G. Philbin, and V. V. Kruglyak, Appl. Phys. Lett. **113**, 212404 (2018).

15. N. Loayza, M. B. Jungfleisch, A. Hoffmann, M. Bailleul, and V. Vlaminck, Phys. Rev. B **98**, 144430 (2018).

16. M. Vogel, B. Hillebrands, and G. von Freymann, arXiv:1906.02301 (2019).

17. H. Yu, O. d' Allivy Kelly, V. Cros, R. Bernard, P. Bortolotti, A. Anane, F. Brandl, F. Heimbach, and D. Grundler, Nat. Commun. **7**, 11255 (2016).

18. C. Liu, J. Chen, T. Liu, F. Heimbach, H. Yu, Y. Xiao, J. Hu, M. Liu, H. Chang, T.Stueckler, S. Tu, Y. Zhang, Y. Zhang, P. Gao, Z. Liao, D. Yu, K. Xia, N. Lei, W. Zhao, and M. Wu, Nat. Commun. **9**, 738 (2018).

19. M. Madami, Y. Khivintsev, G. Gubbiotti, G. Dudko, A. Kozhevnikov, V. Sakharov, A. Stal'makhov, A. Khitun, and Y. Filimonov, Appl. Phys. Lett. **113**, 152403 (2018).

20. E. Albisetti, S. Tacchi, R. Silvani, G. Scaramuzzi, S. Finizio, S. Wintz, J. Raabe, G. Carlotti, R. Bertacco, E. Riedo, and D. Petti, arXiv:1902.09420 (2019).

21. F. Heussner, G. Talmelli, M. Geilen, B. Heinz, T. Brächer, T. Meyer, F. Ciubotaru, C. Adelmann, K. Yamamoto, A. A. Serga, B. Hillebrands, and P. Pirro, arXiv:1904.12744 (2019).

22. V. E. Demidov, M. P. Kostylev, K. Rott, P. Krzysteczko, G. Reiss, and S. O. Demokritov, Phys. Rev. B **83**, 054408 (2011).

23. V. E. Demidov, S. O. Demokritov, K. Rott, P. Krzysteczko, and G. Reiss, Appl. Phys. Lett. **91**, 252504 (2007).

24. V. E. Demidov, S. O. Demokritov, K. Rott, P. Krzysteczko, and G. Reiss, Phys. Rev. B **77**, 064406 (2008).

25. W. B. Case, E. Sadurni, and W. P. Schleich, Opt. Express **20**, 27253 (2012).

26. G. Vitrant, S. Zaiba, B. Y. Vineeth, T. Kouriba, O. Ziane, O. Stéphan, J. Bosson, and P. L. Baldeck, Opt. Express **20**, 26542 (2012).

27. D. Weisman, S. Fu, M. Gonçalves, L. Shemer, J. Zhou, W. P. Schleich, and A. Arie, Phys. Rev. Lett. **118**, 154301 (2017).

28. E. Hecht, Optics. 5th edition. (Pearson, Harlow, 2017).

29. The similarity between a narrow (narrower than half of the wavelength) strip antenna and one-dimensional slit follows from the Huygens–Fresnel principle. In both cases, the appearing patterns can be considered as a result of the interference of secondary wavelets radiated by point sources located on a straight wavefront of finite length.

30. We note that utilization of excitation structures based on coplanar lines with smoothly changing geometrical parameters can help to improve the overall microwave-to-spin wave conversion efficiency. Such structures have been considered in P. Gruszecki, M. Kasprzak, A. E. Serebryannikov, M. Krawczyk, and W. Śmigaj, Sci. Rep. **6**, 22367 (2016) and H. S. Körner, J. Stigloher, and C. H. Back, Phys. Rev. B **96**, 100401(R) (2017).

31. B. A. Kalinikos, IEE Proc. H **127**, 4 (1980).

32. A. Vansteenkiste, J. Leliaert, M. Dvornik, M. Helsen, F. Garcia-Sanchez, and B. Van Waeyenberge, AIP Advances **4**, 107133 (2014).

33. The calculations were performed at $f$=3.8 GHz. The static magnetic field was increased to $H$=2900 Oe to obtain the same spin-wave wavelength of 0.6 μm, as in the case of the in-plane magnetized film.

34. V. E. Demidov, S. O. Demokritov, D. Birt, B. O'Gorman, M. Tsoi, and X. Li, Phys. Rev. B **80**, 014429 (2009).

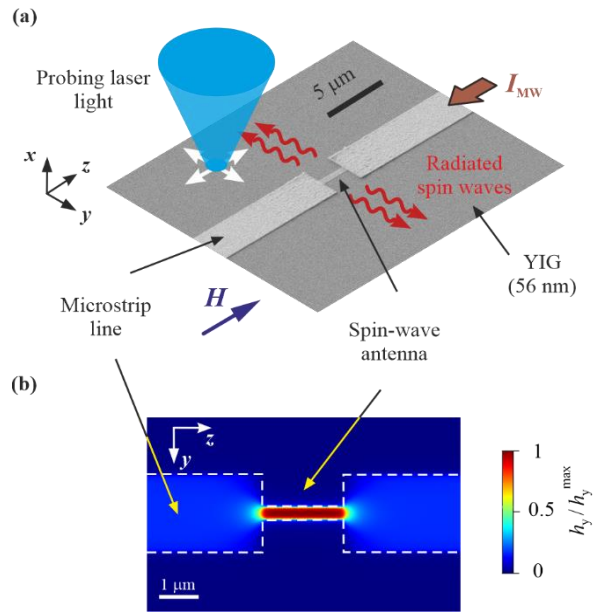

**Figure 1.** (a) Schematic of the experiment. (b) Normalized calculated spatial distribution of the amplitude of the dynamic magnetic field created by the antenna in the YIG film.

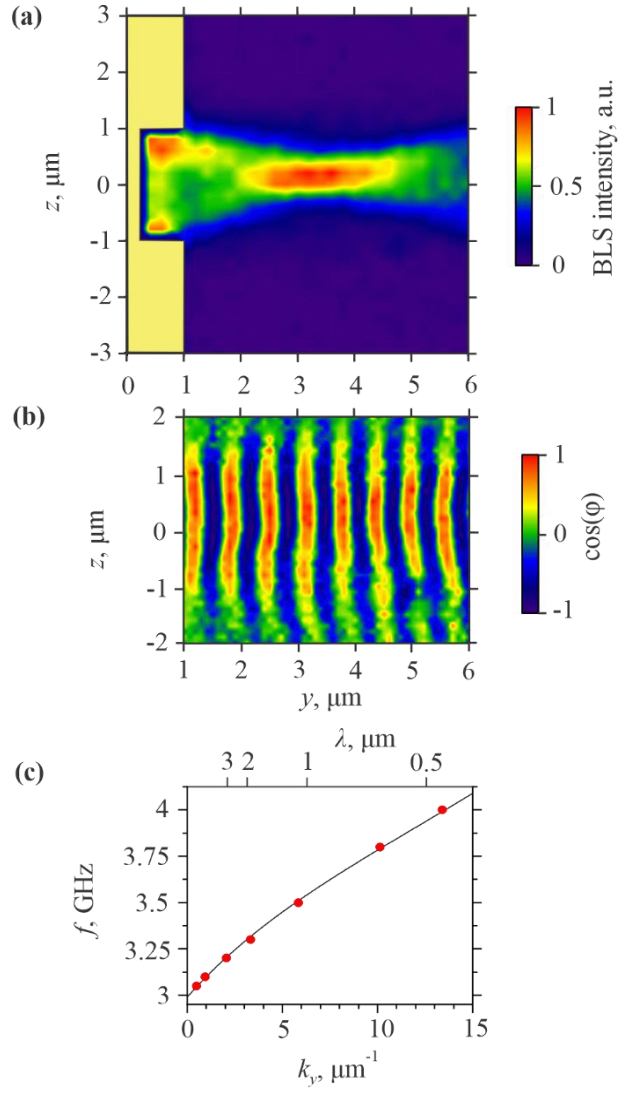

**Figure 2.** Representative spatial maps of the intensity (a) and phase (b) of radiated spin waves recorded by BLS at $H$=500 Oe and $f$=3.8 GHz. (c) Measured (symbols) and calculated (solid curve) dispersion curve of spin waves.

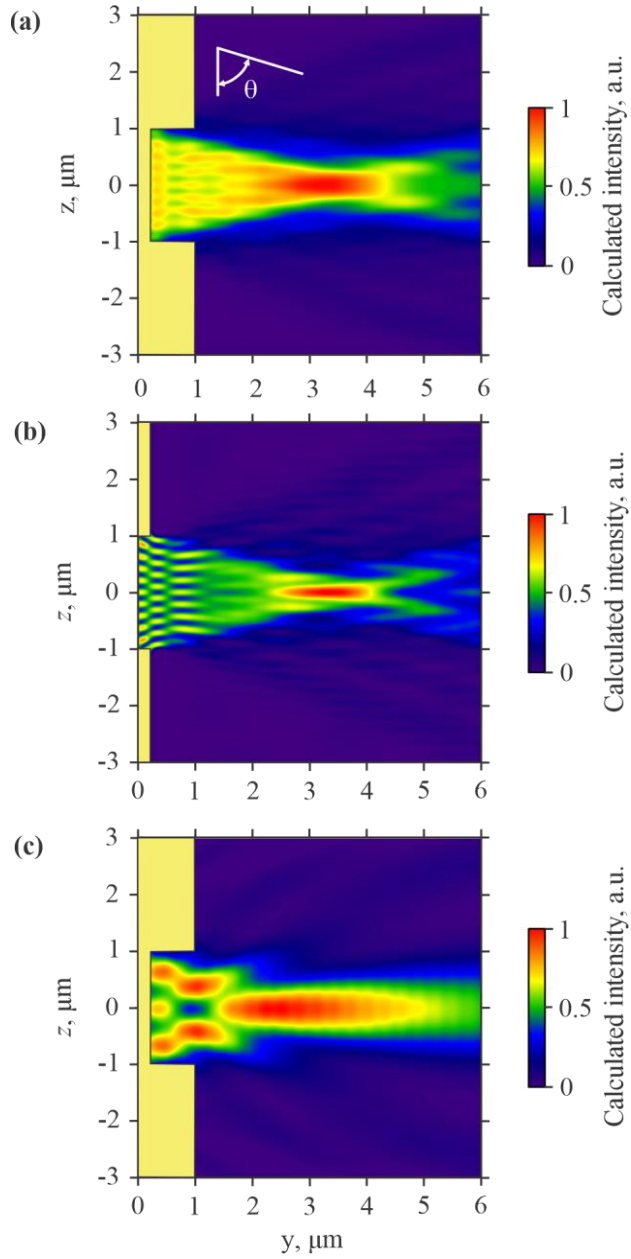

**Figure 3.** (a) Calculated intensity map of spin waves radiated by the antenna. The schematic of the antenna and the connecting microwave lines is superimposed on the map. (b) Calculated intensity map of spin-wave diffraction from a one-dimensional slit. Superimposed rectangles mark the regions with an increased damping. (c) Similar to (a) calculated for the case of isotropic spin-wave dispersion. Calculations were performed for $H$=500 Oe and $f$=3.8 GHz.

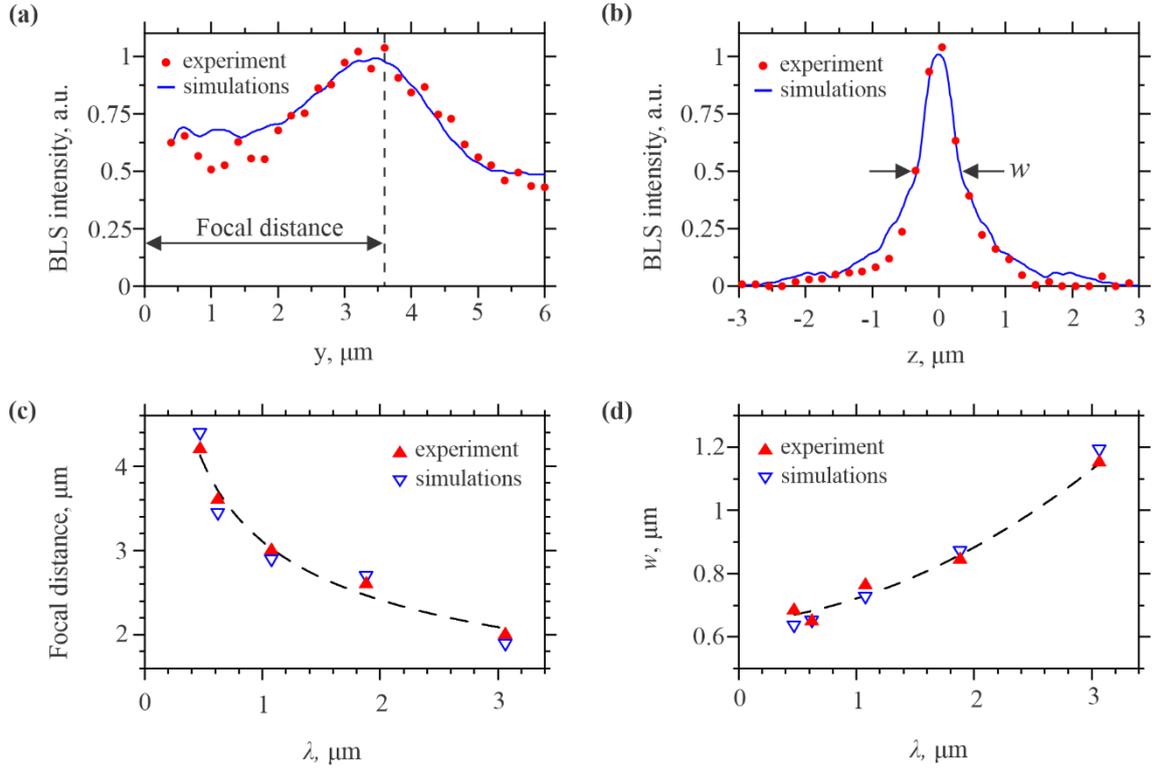

**Figure 4.** (a) One-dimensional sections of the experimental (symbols) and the calculated (solid curve) intensity maps along the axis of the spin-wave beam at $z=0$. (b) Transverse sections of the experimental (symbols) and the calculated (solid curve) intensity maps at the $y$-position corresponding to the maximum intensity. (c) Dependence of the focal distance on the wavelength. (d) Dependence of the width of the spin-wave beam at the focal position on the wavelength. Curves in (c) and (d) are guides for the eye.

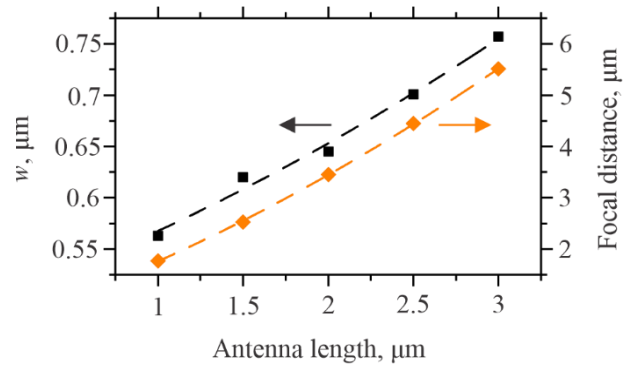

**Figure 5.** Dependences of the beam width at the focal position (squares) and of the focal distance (diamonds) on the length of the antenna calculated for spin waves with the wavelength of 0.6 μm. Dashed curves are guides for the eye.